\documentclass[12pt]{article}
\usepackage{graphicx}
\usepackage{amssymb}
\usepackage{amscd}
\usepackage{amsmath}
\usepackage{appendix}

\begin{document}
\begin{titlepage}

{\hbox to\hsize{\hfill February 2016 }}

\bigskip \vspace{3\baselineskip}

\begin{center}
{\bf \large 
Solving the Strong CP Problem with High-Colour Quarks \\ and Composite Axion }

\bigskip

\bigskip

{\bf Archil Kobakhidze \\ }

\smallskip

{ \small \it
ARC Centre of Excellence for Particle Physics at the Terascale, \\
School of Physics, The University of Sydney, NSW 2006, Australia \\
E-mail: archil.kobakhidze@sydney.edu.au
\\}

\bigskip
 
\bigskip

\bigskip

{\large \bf Abstract}

\end{center}
\noindent 

I propose a new axionic solution to the strong CP problem which involves a hypothetical vector-like quark(s) in a high-colour representation of the conventional QCD.  There are two distinct scenarios. If the current mass of the exotic quark is zero, the strong CP phase can be trivially rotated away. The high-colour quark is `hidden' in various  bounds states, the lightest being the composite axion field, with properties similar to the standard invisible axion. If the high-colour quark acquire a non-zero current mass due to the spontaneous chiral symmetry breaking, the composite axion can be heavy, while the strong CP phase is still cancelling out in the vacuum. The phenomenology and cosmological implications of the heavy composite axion differs drastically from the standard invisible axion. 
 
\end{titlepage}

\vspace{1cm}

\section{The strong CP problem.}
Topologically distinct vacua in quantum chromodynamics (QCD), connected by instantons, imply violation of combined charge conjugation (C) and spatial parity (P) CP symmetry in strong interactions, in addition to the well-established  CP violation in weak interactions. This strong CP violation is parametrised through the strong CP phase $\theta$, which theoretically can resume an arbitrary value, $\theta \in [-\pi, \pi)$. In view of the observed CP violation in weak interactions, the natural theoretical expectation is $\theta \sim {\cal O}(1)$. However, contrary to this expectation,  $\theta$ is constrained to be extremely small, $|\theta|<10^{-10}$, from non-observation of the electric dipole moment for the neutron. This constitutes the long-standing problem in particle physics, known as the strong CP problem, see Ref. \cite{Kim:2008hd} for a recent review. 

Three qualitatively different solutions to the strong CP problem have been proposed in the past. One, arguably the most speculative, is based on trivialising the topology of QCD vacua by assuming the existence of extra spatial dimensions \cite{Khlebnikov:1987zg, Chaichian:2001nx}. The second solution postulates CP invariance for the whole theory, except the vacuum state. In this scenario $\theta=0$ only in the classical limit. To satisfy the experimental bound $|\theta|<10^{-10}$, one must, in general, insure that $\theta$ vanishes also at the 1-loop level. This makes the models that incorporate this scenario rather complicated and contrived \cite{Beg:1978mt, Georgi:1978xz, Nelson:1983zb, Barr:1984qx}.  Most importantly, the experimental data is in excellent agreement with the Cabibbo-Kobayashi-Maskawa  mechanism, where CP is explicitly, not spontaneously broken in the weak sector of the theory, so many models of this category are in fact excluded. 

The third solution to the strong CP problem is arguably the most economic and elegant. It postulates an extra U(1) global chiral symmetry, which within just Standard Model can be achieved assuming at least one quark flavour is massless, i.e. the current mass for up quark is $m_u=0$. Unfortunately, the lattice calculations seem to exclude massless quarks \cite{Nelson:2003tb}.  Beyond the Standard Model, one can assume that a chiral U(1) is spontaneously broken, resulting in a pseudo-Goldstone boson, known as the axion. It has been shown first by Peccei and Quinn  that the dynamics of the axion field is arranged such that its expectation value cancels out almost exactly the strong $\theta$ phase \cite{Peccei:1977hh} (see also Refs. \cite{Weinberg:1977ma, Wilczek:1977pj} and \cite{Kim:1979if, Shifman:1979if, Zhitnitsky:1980tq, Dine:1981rt}). The experimental bound on the hypothetical axion particle requires very high scale for U(1) symmetry breaking, $\gtrsim 10^9$ GeV \cite{Kim:2008hd}, and hence, in general, this scenario is plagued with the familiar mass hierarchy problem\footnote{Supersymmetry or scale invariance \cite{Foot:2013hna, Kobakhidze:2014afa} may be the cure, see \cite{Volkas:1988cm, Clarke:2015bea}.}. 

In this paper, I propose a new mechanism for solving the strong CP problem, which belongs to the third category. The key postulate behind of this solution is the existence of a vector-like quark(s) in high-colour representation of  the QCD gauge symmetry group SU(3). There is an U(1) symmetry under the chiral rotation of this quark, which is dynamically broken at some high energy scale $f_a>>f_{\pi}\approx 130$ MeV [$f_{\pi}$ being the familiar pion decay constant] by a high-colour quark - antiquark condensate. This breaking is accompanied by a composite pseudo-Goldstone particle, the axion. There are two scenarios to be distinguished. If the exotic quark is massless, the strong CP phase $\theta$ can be trivially rotated away, just like in the case with $m_u=0$. In this case, the axion particle composed of high-colour quark - antiquark pair is very light, similar to the standard invisible axion  \cite{Kim:1979if, Shifman:1979if, Zhitnitsky:1980tq, Dine:1981rt}. Another possibility is when the exotic quark acquires current mass via Yukawa coupling to a complex scalar field. The axion in this case  can be heavy, while the strong CP phase still cancels out in the vacuum. The phenomenology of the composite axion may thus differ drastically from the standard axion.  Besides being stunningly economical, the proposed scenario is also free from the hierarchy problem, since the mass scales are generated dynamically due to the high-colour quark condensation. 

Before proceeding further, I would like to acknowledge previous works which have some relevance to my proposal. To ameliorate the hierarchy problem, the models of composite invisible axion have been constructed in \cite{Kim:1984pt, Kaplan:1985dv, Randall:1992ut} based on additional confining interactions with higher than QCD confinement scale. Similar hypothetical strong dymamics can be used to make the standard axion heavier \cite{Rubakov:1997vp, Berezhiani:2000gh, Hook:2014cda, Fukuda:2015ana}. In \cite{Tye:1981zy, Holdom:1982ex} it was suggested that if QCD itself becomes stronger at high energy scales, the small-size instanton contribution can enhance the mass of the standard axion. Finally, while I was preparing this paper, the work \cite{Wilczek:2016gzx} appeared on arXiv. The authors propose the solution to the strong CP problem by introducing massless exotic quarks, which are confined by an extra QCD-like forces into heavy exotic hadrons. In the second version of the paper they have also correctly identified the light composite axion state.    

\section{The composite axion}
I will start by discussing the properties of the composite axion that emerge in the low energy limit of the theory first. In the next section, I will turn to the two specific scenarios that solves the strong CP problem and gives rise of such axion. Let me postulate the existence of a vector-like quark $Q$ in some $R$ (other than triplet) representation of the QCD SU(3) group\footnote{High-colour quarks have a long history, see, e.g. Refs \cite{Ma:1975qy, Karl:1976jk, Wilczek:1976qi, Ng:1978qt, Georgi:1979ng, Marciano:1980zf} for the earliest papers. In Ref. \cite{Marciano:1980zf} it was suggested that quarks in sextet colour representation and also carrying the appropriate eletroweak quantum numbers may break the electroweak symmetry dynamically. My motivation for introducing the high-colour quark is entirely different.}. Suppose this quark carries a weak hypercharge $y_Q$ and has a current mass $m_Q$, which supposedly is generated from the Yukawa interactions with an extra complex scalar field (see, the next section ).  Other assignment of electroweak quantum numbers are also possible, but I stick to this simple one here.  I assume that $m_Q<< f_a$ and hence identify an approximate $U(1)_A^Q$ symmetry of chiral rotations of the $Q$ quark. This is an extra axial symmetry required for the solution of the strong CP problem. $U(1)_A^Q$ symmetry is broken explicitly by the current mass $m_Q$ and the colour and hypercharge anomalies.   The associated current density and its divergence read:
\begin{eqnarray}
\label{1}
 J_A^{\mu}&=&\bar Q \gamma^{\mu}\gamma_5 Q,~  \\
\partial_{\mu}J_A^{\mu}&=& 2im_Q\bar Q\gamma_5 Q+
\frac{T(R)\alpha_3}{4\pi}\epsilon^{\mu\nu\alpha\beta}G_{\mu\nu}^aG_{\alpha\beta}^a
+\frac{d(R) y_{Q}^2\alpha_1}{4\pi}\epsilon^{\mu\nu\alpha\beta}B_{\mu\nu}B_{\alpha\beta}~,
\label{2}
\end{eqnarray}  
where $G_{\mu\nu}^a$ and $B_{\mu\nu}$ are gluon and hypercharge field strength tensors, respectively; $\alpha_3$ is the colour SU(3) and $\alpha_1$ is the U(1)$_Y$ hypercharge fine-stucture constants; $T(R)$ is the Dynkin index for the SU(3) representation $R$, which is expressed through the corresponding eigenvalue of the quadratic Casimir operator, $C_2(R)$, and the dimension of the representation, $d(R)$, as:
\begin{equation}
T(R)=\frac{1}{8}C_{2}(R)d(R)~. 
\label{3}
\end{equation}

As to the conventional quarks, it is suffice to consider only two flavours of the lightest colour-triplet quarks, $q=(u, d)^{\rm T}$.  The relevant axial symmetries are $U(1)_A^{q}$, generated by an overall chiral phase transformation of $q$, and the Cartan subgroup of the axial flavour $SU(2)_A$. Both of this symmetries are broken explicitly by the quark non-zero masses $m_{u,d}$. $U(1)_A^{q}$ in addition is broken by the hypercharge and colour anomalies. The corresponding currents and their divergences are: 
\begin{eqnarray}
\label{4} 
j_A^{\mu}&=&\bar q \gamma^{\mu}\gamma_5 q,~ j_3^{\mu}=\bar q \sigma^3\gamma^{\mu}\gamma_5 q  \\
\partial_{\mu}j_A^{\mu}&=& 2i\sum_{i={u,d}}m_i\bar q_i\gamma_5 q_i+ 
\frac{\alpha_3}{4\pi}\epsilon^{\mu\nu\alpha\beta}G_{\mu\nu}^aG_{\alpha\beta}^a
+\frac{3(y_u^2+y_d^2)\alpha_1}{8\pi}\epsilon^{\mu\nu\alpha\beta}B_{\mu\nu}B_{\alpha\beta}~, 
\label{5} \\ 
\partial_{\mu}j_3^{\mu}&=& 2i(m_u\bar u\gamma_5 u-m_d\bar d\gamma_5 d)+ 
+\frac{3(y_u^2-y_d^2)\alpha_1}{8\pi}\epsilon^{\mu\nu\alpha\beta}B_{\mu\nu}B_{\alpha\beta}~,
\label{6}
\end{eqnarray} 
where $\sigma^{3}=diag(1,-1)$ is the third Pauli matrix, acting in the flavour space; $y_u=4/3$ and $y_d=-2/3$ are hypercharges of the weak isospin singlet up, $u_R$, and the isospin down, $d_R$, quarks,  respectively. 
 
The axial $U(1)_{A}^q$ and $U(1)_A^{Q}$ are also broken spontaneously respectively by the quark condensates:
\begin{eqnarray}
\label{7}
\langle 0 \vert \bar q q\vert 0\rangle \approx - c_qf_{\pi}^3~, \\
\langle 0 \vert \bar Q Q\vert 0\rangle \approx - c_Qf_a^3~,
\label{8}
\end{eqnarray}  
where $c_q, c_Q$ are constants $\sim \mathcal{O}(1)$ and  $\vert 0\rangle$ denotes the vacuum state. Due to the high-colour representation, the scale of axial symmetry breaking in $Q$-sector is hierarchically larger than the scale of the chiral symmetry breaking in the conventional quark sector, $f_a >> f_{\pi}\approx 130$ MeV. To estimate $f_a$, we follow \cite{Marciano:1980zf} and assume that the chiral symmetry breaking determined by the strength of the quark-antiquark binding potential and occurs at some critical value of the QCD string tension which is common for both $q$ and $Q$ sectors. Then one reasonably expects that
\begin{equation}
C_2(3)\alpha_3({\Lambda_q})=C_2(R)\alpha_3({\Lambda_Q})~,
\label{8a}
\end{equation} 
and $f_a/f_{\pi}\approx \Lambda_{Q}/\Lambda_{q}$, where $\Lambda_q$ and $\Lambda_Q$ are typical binding energy scales in $q$ and $Q$ sectors, respectively. Utilising the relevant renormalization group equation (RGE), one can solve Eq. (8a) for $\Lambda_{Q}/\Lambda_{q}$. In a crude approximation, taking 1-loop RGE with 6 flavours of ordinary quarks one obtains \cite{Marciano:1980zf}:   
\begin{equation}
f_a\approx f_{\pi} \exp\left[\frac{2\pi}{7\alpha_3(\Lambda_q)}\left(\frac{3}{4}{\rm C}_2(\mathcal{R})-1\right)\right]~,
\label{8b}
\end{equation}
where $\Lambda_{q}\sim 1$ GeV. One can see that $f_a$ is larger than $f_{\pi}$ by an exponential factor, which is very sensitive to values of the strong coupling constant $\alpha_3$ at low energies. At the same time, determination of $\alpha_3$ at low energies contains significant uncertainties. For example, the estimate obtained from the charmonium fine stricture splitting is $\alpha_3(1~{\rm  GeV})\approx 0.38\pm 0.05$ \cite{Badalian:1999fq}, while extracting $\alpha_3$ from the hadronic decays of tau lepton gives $\alpha_3(1.7~{\rm GeV})\approx 0.331\pm 0.013$ \cite{Pich:2013lsa}. In our estimations below we allow $\alpha_3(\Lambda_{q})=0.3-0.5$. To have a feel of numbers, I presented some estimates in Table 1.
\begin{table}
\begin{center}
\begin{tabular}{|l|l|l|l|}
\hline
Repr. & C$_2(R)$ & T($R$) & $f_a$, GeV \\
\hline
\hline
d($R$)=6 & $10/3$ & $5/2$ & $2.0-12.0$ \\
\hline
d($R$)=8 &  $3$ & $3$ & $1.0-6.0$ \\
\hline
d($R$)=10 & $6$&  $15/2$ & $75.0-4590.0$ \\
\hline
d($R$)=15 &  $28/3$ & $35/2$ &$(0.007-8.0)\cdot 10^6$ \\
\hline
d($R$)=21 & $40/3$ & $35 $ &$(10^{-4}-5.0)\cdot 10^{11}$ \\
\hline
\end{tabular}
\end{center}
\caption{\small Estimates of $f_a$ for various high-colour representations according to Eq. (\ref{8b}). The strong coupling is assumed in the range $\alpha_3 (\Lambda_{q})=0.3-0.5$. }
\end{table}

The physical (mass eigenstate) composite axion, which is essentially a bound state of exotic quark-antiquark pair, $a\sim \bar Q\gamma_5 Q$, couples to the colour anomaly free combination of the axial currents (\ref{1}) and (\ref{4}):      
\begin{equation}
\tilde J^{\mu} =J_A^{\mu}- T(R)\left( j^{\mu}_A + \kappa j_3^{\mu} \right)~,
\label{9}
\end{equation}
where $\kappa = \frac{m_d-m_u}{m_d+m_u}\approx 0.35$. The corresponding charge is:
\begin{equation}
\tilde Q_5=\int d^3x~\tilde J^{0}
\label{10}
\end{equation}

The mass of the composite axion can be computed using Dashen's formula \cite{Dashen:1969eg}: 
\begin{eqnarray}
m_a^2 &\simeq& -\frac{1}{f_a^2}\langle 0 \vert [\tilde Q_5, \partial_{\mu}\tilde J^{\mu}]\vert 0 \rangle 
\approx -\frac{4}{f^2_a}\left( m_{Q}\langle 0 \vert\bar QQ\vert 0 \rangle +T^2(R)\frac{m_um_d}{m_u+m_d}\langle 0 \vert\bar qq\vert 0 \rangle\right) \nonumber \\
&=&4c_Q m_Qf_a + 4c_qT^2(R)\left(\frac{f_{\pi}}{f_a}\right)^2\frac{m_um_d}{m_u+m_d}f_{\pi} ~,
\label{11}
\end{eqnarray} 
Being a true pseudo-Goldstone boson, the mass of the composite axion vanishes as $m_Q\to 0$ and $m_u({\rm or}~ m_d)\to 0$.  

The axion-hypercharge coupling also can be readily read off from the hypercharge anomaly of $\tilde J^{\mu}$ (\ref{9}):
\begin{eqnarray}
\mathcal{L}&\supseteq& C_{aBB}\epsilon^{\mu\nu\alpha\beta}a B_{\mu\nu}B_{\alpha\beta}~, \nonumber \\
C_{aBB}&=&\frac{\alpha_1}{4\pi}\frac{1}{f_a}\left(d(R)y_Q^2-\frac{2T(R)}{3}\frac{4m_d+m_u}{m_d+m_u}\right) 
\label{12}
\end{eqnarray}

The coupling of the composite axion to ordinary matter (composed of quarks $q$) as well as exotic composite bound states of $Q$ quarks  is also determined by saturating the non-anomalous axial current (\ref{9}) Ward identity and is characterized by the Yukawa couplings:
 \begin{eqnarray}
\mathcal{L}\supseteq \frac{2i}{f_a}\left(m_Q \bar Q\gamma_5 Q + T(R)\frac{m_um_d}{m_u+m_d} \bar q\gamma_5 q \right)~.
\label{12a}
\end{eqnarray}
One observes a scaling $\sim m/f_a$, which is typical for Yukawa couplings of an axion.

\section{Cancelling the strong CP phase: light and heavy axions.} I consider now more explicitly two distinct cases $m_Q=0$ and $m_Q\neq 0$.  
\subsection{$m_Q=0$} 
The solution to the strong CP problem with the above composite axion is standard. Let imagine first that the exotic quark $Q$ is massless, $m_Q=0$. This implies that the classical Lagrangian is exactly invariant under $U(1)_A^Q$ chiral rotation of $Q$, while the corresponding measure $\mathcal{D}\bar Q \mathcal{D}Q$ in the generating functional is not, due to the colour anomaly. Hence, we can perform $U(1)_A^Q$ rotation of $Q$ such that to exactly cancel out the strong CP phase, $\sim \theta G\tilde G$. The strong CP phase is not physical in this case and, in accord with observations, CP is a good symmetry of strong interactions. The vacuum expectation value of the composite axion relaxes to zero. 

For vanishing $m_Q$, the mass of the composite axion is generated through the dynamics of light ordinary quarks, as it is in fact the case for the standard axion. The axion is light in this case: 
\begin{equation}
\left. m_a \right \vert_{m_Q=0}\simeq 2T(R)m_{\pi}\frac{f_{\pi}}{f_a}\frac{\sqrt{m_um_d}}{m_u+m_d} 
=\left \lbrace 
\begin{tabular}{ll}
123.0 $\left(\frac{f_a}{1~{\rm TeV}}\right)^{-1}$ KeV,  &  for $d(R)=10$ \\
29.0 $\left(\frac{f_a}{10~ {\rm TeV}}\right)^{-1}$ KeV, &   for $d(R)=15$\\
6.0 $\left(\frac{f_a}{10^{8}~ {\rm TeV}}\right)^{-1}$ meV, &   for $d(R)=21$ \\
\end{tabular}
\right. ~,
\label{13}
\end{equation}     
where $m_{\pi}=\sqrt{c_q(m_u+m_d)f_{\pi}}\approx 135$ MeV is the pion mass. Eq. (\ref{13}) coincides with the mass formula for the standard axion up to a factor $2T(R)$ which reflects an enhancement due to the high-colour representation of the constituent quark $Q$. In view of couplings to ordinary matter (\ref{12a}), $10$- and $15$-plets are excluded by observations, while the mass and couplings of an axion composed of 21-plets is in typical  invisible axion ballpark. Note, however, that this is a `passive' axion, which relaxes to zero in the vacuum, since the strong CP phase has been removed by chiral transformations of $Q$. 

\subsection{$m_Q\neq 0$} 
To generate the current mass for $Q$ quark, such that the perturbative vacuum still maintains $U(1)_A^{Q}$ degeneracy, I introduce the complex scalar field $\phi$ and its   $U(1)_A^{Q}$-symmetric interaction with $Q$ quark:
\begin{equation}
\mathcal{L_{\phi}}\supset    \left\vert \partial_{\mu}\phi\right\vert^2-m_{\phi}^2\vert\phi\vert^2- \lambda_Q\bar Q_LQ_R\phi +h.c.+...~,  
\label{14}
\end{equation} 
where $m_{\phi}^2>0$ and I have omitted the scalar self-interaction terms under assumption that they are small.  
The $\bar QQ$ condensation (\ref{8}) triggers condensation of the scalar field, 
\begin{equation}
\langle 0\vert  \phi \vert 0\rangle\simeq -\frac{\lambda_Q}{m_{\phi}^2}\langle 0\vert \bar Q Q\vert 0 \rangle~. 
\label{15}
\end{equation}
In this vacuum $Q$ quark develops the current mass, 
\begin{equation}
m_Q= -\frac{\lambda^2_Q}{m_{\phi}^2}\langle 0\vert \bar Q Q\vert 0 \rangle~. 
\label{16}
\end{equation}
 For sufficiently large $m_Q$ the dominant contribution to the axion mass comes from the first term on the second line of Eq. (\ref{11})\footnote{The composite axion is mixed with the heavy ($\sim m_{\phi}^2$) pseudo-scalar from $\phi$ as well with light ordinary mesons $\pi^0$ and $\eta$. I assume here that these mixings are small.}:
\begin{equation}
m_a \simeq 2c^{1/2}_Q \sqrt{m_Qf_a} ~. 
\label{17}
\end{equation}
It is clear that, depending on $m_Q$ and $f_a$, the composite axion can, in principle, be quite heavy,  $m_a \lesssim f_a$. Although $d(R)=6,8$ cases are most likely excluded due to the unacceptably low $f_a$, $d(R)=10$ (and higher), with $f_a\gtrsim $TeV, represents  a viable option.      

The heavy composite axion acts just like the standard axion. Indeed, since $U(1)_A^{Q}$ symmetry is explicitly broken only by anomalies,  the strong CP phase can be absorbed by shifting the axion field $a(x)$, which then relaxes to the trivial ground state, cancelling out unacceptably large CP violation in the QCD sector. The interesting twist here is that, unlike the standard invisible axion, the heavy axion can be probed at the LHC or future high-energy colliders, providing the axion mass is within the reach. In fact, my collaborators and I contemplated recently \cite{Barrie:2016ntq} that the heavy composite axion may be responsible for the excess of events in the diphoton channel seen at 750 GeV in the early LHC Run 2 data \cite{ATLAS750, CMS750}. Phenomenological aspects and potential cosmological implications of the heavy axion will be discussed in subsequent publications. 

\section{Conclusion}
In this paper, I proposed a new axionic solution to the strong CP problem by hypothesising a vector-like quark(s) in a high-colour representation of the ordinary QCD. The chiral symmetry associated with the exotic quark is broken by the quark-antiquark condensate at a scale $\sim f_a$, much higher than the scale of the chiral symmetry breaking, $\sim f_{\pi}$, in the sector of conventional triplet quarks, resulting in a composite axion. There are two distinct cases. If the current mass of the exotic quark is zero, the composite axion is light is very similar to the standard axion. The instanton transitions are absent in this case and, hence, the strong CP phase, being an unphysical  parameter is trivially rotated away via chiral phase transformation of the exotic quark. In another scenario, I assume that non-zero current mass of the exotic quark is generated through its coupling with the complex scalar field. Depending on this mass and $f_a$, the axion in this case may be heavy, while the standard Peccei-Quinn mechanism for cancellation of the strong CP phase is still operative. Phenomenology and cosmology of such a heavy axion differs dramatically from the standard invisible axion. The heavy axion may be accessible in high-energy collider experiments. It may also contribute to the dynamical generation of matter-antimatter asymmetry through the anomalous coupling to the hypercharge field. Finally, the exotic quark may form the stable bound states that could serve as dark matter. These aspects of the current proposal deserve further study.

\paragraph{Acknowledgement.} I am indebted to my CoEPP colleagues for discussion during the 2016 CoEPP Annual Workshop. This work was partially supported by the Australian Research Council and Shota Rustaveli National Science Foundation (grant DI/12/6-200/13).

\end{document}